\documentclass[aps,prb,twocolumn,superscriptaddress,floatfix]{revtex4-2}

\usepackage{amsmath,amssymb,amsfonts}
\usepackage{bm}
\usepackage{mathtools}
\usepackage{graphicx}
\usepackage{tabularx}
\usepackage{xcolor}
\usepackage{booktabs}
\usepackage{array}
\usepackage{multirow}
\usepackage{lineno}
\usepackage[ruled,vlined,linesnumbered]{algorithm2e}
\usepackage[caption=false]{subfig}
\usepackage{ragged2e}
\usepackage{braket}

\usepackage{placeins}
\usepackage[colorlinks=true,linkcolor=blue,citecolor=blue,urlcolor=blue]{hyperref}

\usepackage{tikz}
\usetikzlibrary{
  arrows.meta,
  positioning,
  shapes.geometric,
  calc
}

\begin{document}
\author{Bishal Thapa}
\affiliation{Department of Physics and Astronomy, George Mason University, Fairfax, VA 22030, USA}
\affiliation{Quantum Science and Engineering Center, George Mason University, Fairfax, VA 22030, USA}

\author{Phanish Suryanarayana}
\affiliation{College of Engineering, Georgia Institute of Technology, Atlanta, GA 30332, USA}
\affiliation{College of Computing, Georgia Institute of Technology, Atlanta, GA 30332, USA}
\author{Igor I. Mazin}
\email[Email: ]{imazin2@gmu.edu}
\affiliation{Department of Physics and Astronomy, George Mason University, Fairfax, VA 22030, USA}
\affiliation{Quantum Science and Engineering Center, George Mason University, Fairfax, VA 22030, USA}

\title{Assessing the suitability of the Thomas-Fermi-von Weizs\"acker density functional for itinerant magnetism}
\date{\today }

\begin{abstract}
We assess the ability of the Thomas--Fermi--von Weizs\"acker (TFW) functional within orbital-free density functional theory (DFT) to describe itinerant magnetism. Magnetic stability is evaluated through the susceptibility obtained from the second derivative of the total energy with respect to the net magnetization. Calculations are performed for the paramagnetic metals Al and Pd and the canonical ferromagnets Fe, Co, and Ni, with the results benchmarked against Kohn--Sham DFT. The orbital-free results show poor agreement with the Kohn--Sham predictions, failing to capture even the qualitative trends. Using the orbital-free ground-state density with the Kohn--Sham functional in a non-self-consistent calculation yields reasonable qualitative agreement, although the quantitative agreement remains limited. These results highlight fundamental limitations of the TFW functional for describing itinerant magnetism.
\end{abstract}
\maketitle


\section{Introduction}
Orbital-free density functional theory (OF-DFT) \cite{doi:10.1021/acs.chemrev.2c00758} and Kohn--Sham DFT \cite{PhysRev.140.A1133} are two practical realizations of the formally exact DFT framework \cite{PhysRev.136.B864}, differing primarily in how the non-interacting kinetic energy is evaluated. In Kohn--Sham DFT, the interacting many-electron system is mapped onto an auxiliary system of non-interacting electrons whose density is constructed from single-particle orbitals, allowing the kinetic energy to be computed exactly from these orbitals. In contrast, orbital-free DFT eliminates the explicit orbitals and instead approximates the kinetic energy directly as a functional of the electron density.

The accuracy of orbital-free DFT therefore hinges on the availability of reliable approximations for the non-interacting kinetic energy as a functional of the electron density. Early examples include the Thomas–Fermi \cite{Thomas_1927,PhysRev.75.1561} and Thomas–Fermi–von~Weizs{\"a}cker (TFW) \cite{benguria1981thomas} functionals. Subsequent developments introduced a range of local and semilocal forms incorporating higher-order derivatives of the density \cite{PhysRevA.20.586,PhysRevB.88.161108,PhysRevB.98.041111,PhysRevB.100.165111,Laricchia2011,PhysRevB.99.155137}. Nonlocal kinetic-energy functionals derived from linear-response theory have also been developed, including the Chac{\'o}n–Alvarellos–Tarazona \cite{PhysRevB.32.7868,PhysRevB.53.9509,PhysRevB.57.4857}, Wang–Teter \cite{PhysRevB.45.13196}, and Wang–Govind–Carter \cite{PhysRevB.60.16350} functionals.

A key advantage of orbital-free DFT relative to the Kohn-Sham variant is its computational efficiency. In Kohn–Sham DFT, the solution of the eigenvalue problem for the occupied states leads to a computational cost that scales cubically with system size. Orbital-free DFT avoids the explicit computation of orbitals and therefore exhibits linear scaling with system size, enabling efficient simulations of large systems containing millions of atoms \cite{doi:10.1021/acs.chemrev.2c00758}. This efficiency, however, comes with important limitations. In particular, since orbital-free DFT does not compute Kohn–Sham eigenstates, it does not provide access to the electronic spectrum and therefore lacks direct information about the electronic density of states.

The absence of explicit electronic states is often viewed as a major limitation when considering itinerant magnetism. In band-theory descriptions of magnetic materials, magnetic instabilities arise from the competition between the kinetic-energy cost of spin polarization and the exchange-correlation energy gain. Within Kohn--Sham DFT, this balance is closely tied to the electronic structure near the Fermi level and to quantities derived from the density of states. In principle, orbital-free DFT can be generalized to treat magnetic systems by introducing spin-resolved densities. Spin-polarized extensions of Thomas--Fermi-type theories have been discussed previously in the literature \cite{oliver1979spin,gunnarsson1976band,janas2023enhancing}, although they have received relatively limited attention.

One reason for this limited attention is that the electronic-structure features responsible for magnetic instabilities, particularly in transition metals, are closely tied to the detailed structure of the density of states, which orbital-free DFT does not explicitly resolve. Nevertheless, it remains of interest to examine how spin-polarized orbital-free formulations perform across different materials. Earlier attempts to extend orbital-free DFT to magnetic systems, including studies of spin-polarized molecules, have highlighted the challenges of describing spin polarization within orbital-free frameworks \cite{xia2012can}. Understanding the extent to which magnetic response can be captured within orbital-free DFT is therefore important for delineating the domain of applicability of orbital-free approaches and for assessing their ability to describe itinerant magnetism.

In this work, we assess the ability of the TFW functional within orbital-free DFT to describe itinerant magnetism. Magnetic stability is evaluated through the susceptibility obtained from the second derivative of the total energy with respect to the net magnetization. Calculations are performed for the paramagnetic metals Al and Pd and the canonical ferromagnets Fe, Co, and Ni, with results compared to Kohn--Sham DFT. The orbital-free predictions show poor agreement with the Kohn--Sham results, even qualitatively. We therefore also examine an approach in which the Kohn--Sham functional is evaluated non-self-consistently on the orbital-free ground-state densities. This yields reasonable qualitative agreement with the Kohn--Sham results, although quantitative agreement remains limited, highlighting fundamental limitations of the TFW functional for describing itinerant magnetism.


\section{Background}
\label{sec:background}

\subsection{Density Functional Theory}
\label{subsec:dft}

Within spin-polarized DFT, the total energy can be written as a functional of the spin densities $n_{\uparrow}(\mathbf{r})$ and $n_{\downarrow}(\mathbf{r})$:
\begin{align}
E[n_{\uparrow},n_{\downarrow}]
&= T_s[n_{\uparrow},n_{\downarrow}]
+ E_{\mathrm{elec}}[n]
+ E_{\mathrm{xc}}[n_{\uparrow},n_{\downarrow}],
\label{eq:dft_total_energy}
\end{align}
where $T_s[n_{\uparrow},n_{\downarrow}]$ is the non-interacting kinetic energy, $E_{\mathrm{elec}}[n]$ is the total electrostatic energy, $E_{\mathrm{xc}}[n_{\uparrow},n_{\downarrow}]$ is the exchange-correlation energy, and the total electron density is
\begin{align}
n(\mathbf{r}) = n_{\uparrow}(\mathbf{r}) + n_{\downarrow}(\mathbf{r}).
\label{eq:total_density}
\end{align}

In Kohn--Sham DFT, the kinetic energy is evaluated exactly for an auxiliary noninteracting system in terms of the spin-dependent Kohn--Sham orbitals $\psi_{i\sigma}(\mathbf{r})$:
\begin{align}
T_s^{\mathrm{KS}}[n_{\uparrow},n_{\downarrow}]
&= -\frac{1}{2}\sum_{\sigma=\uparrow,\downarrow}\sum_i
\int \psi_{i\sigma}^*(\mathbf{r}) \nabla^2 \psi_{i\sigma}(\mathbf{r}) \, d\mathbf{r},
\label{eq:ks_kinetic_energy}
\end{align}
where the index $i$ runs over the occupied orbitals of spin $\sigma$. Indeed, the orbitals are constrained to reproduce the spin densities:
\begin{align}
n_{\sigma}(\mathbf{r})
&= \sum_i \left| \psi_{i\sigma}(\mathbf{r}) \right|^2.
\label{eq:spin_density_from_orbitals}
\end{align}

In TFW orbital-free DFT, the kinetic energy is approximated directly as a density functional:
\begin{align}
T_s^{\mathrm{TFW}}[n_{\uparrow},n_{\downarrow}]
&= C_{\mathrm{TF}} \sum_{\sigma=\uparrow,\downarrow}
\int n_{\sigma}^{5/3}(\mathbf{r}) \, d\mathbf{r}
\nonumber\\
&\quad +  \frac{\lambda}{8}\sum_{\sigma=\uparrow,\downarrow}
\int \frac{|\nabla n_{\sigma}(\mathbf{r})|^2}{n_{\sigma}(\mathbf{r})} \, d\mathbf{r}.
\label{eq:tfw_kinetic_energy}
\end{align}
where $C_{\mathrm{TF}} = \frac{3}{10}(6\pi^2)^{2/3}$ is the Thomas--Fermi coefficient, and $\lambda$ is a parameter that acts as the prefactor of the von Weizs\"acker correction. The first term is a local approximation to the kinetic energy, while the second term provides a gradient correction.

\subsection{Magnetic stability of the paramagnetic state}
\label{subsec:magnetic_stability}

To assess the stability of the paramagnetic state, we examine how the total energy varies with the total magnetization
\begin{align}
M
&= \int \bigl(n_{\uparrow}(\mathbf{r})-n_{\downarrow}(\mathbf{r})\bigr)\, d\mathbf{r}.
\label{eq:magnetization_definition}
\end{align}
The energy at fixed magnetization is obtained by minimizing the energy functional with respect to the spin densities under this constraint:
\begin{align}
E(M)
&= \min_{n_{\uparrow},\,n_{\downarrow}\rightarrow M}
E[n_{\uparrow},n_{\downarrow}].
\label{eq:energy_fixed_magnetization}
\end{align}
Indeed, the paramagnetic reference state corresponds to $M=0$. 

In the absence of external magnetic fields, expanding the energy about the paramagnetic state gives:
\begin{align}
E(M)
&= E(0)
+ \frac{1}{2}
\left.
\frac{d^2E}{dM^2}
\right|_{M=0}
M^2
+ \mathcal{O}(M^4) ,
\label{eq:energy_expansion_m}
\end{align}
where odd powers of $M$ do not appear because the system is invariant under spin inversion ($M\rightarrow -M$), which implies $E(M)=E(-M)$. The second derivative of the energy with respect to magnetization is the inverse uniform static spin susceptibility, also referred to as the inverse Stoner-enhanced magnetic susceptibility:
\begin{align}
\chi^{-1}
&=
\left.
\frac{d^2E}{dM^2}
\right|_{M=0}.
\label{eq:chi_inverse_definition}
\end{align}

The sign of $\chi^{-1}$ can be used to determine the stability of the paramagnetic state:
\begin{align}
\text{paramagnetic state} =
\begin{cases}
\text{stable} & \chi^{-1} > 0, \\
\text{neutral} & \chi^{-1} = 0, \\
\text{unstable} & \chi^{-1} < 0,
\end{cases}
\label{eq:stability_conditions}
\end{align}
where the unstable case corresponds to a transition from the paramagnetic state to a ferromagnetic state. Within Kohn--Sham DFT, this condition can be recast in terms of the Pauli susceptibility and the Stoner parameter by separating the curvature of the total energy into kinetic and exchange-correlation contributions.  This decomposition provides a convenient way to analyze whether exchange-correlation effects overcome the kinetic-energy cost of spin polarization and thereby destabilize the paramagnetic state. In orbital-free DFT, however, such a separation is not evident, and we therefore examine the curvature of the total energy directly.

\section{Results and discussion}

All calculations are performed using the M-SPARC code \cite{xu2020m, zhang2023version}, a \texttt{Matlab} implementation of the large-scale parallel electronic structure code SPARC \cite{xu2021sparc, zhang2023version}. It employs a real-space finite-difference discretization, whose formulation and implementation for both orbital-free and Kohn--Sham DFT have been described previously \cite{ghosh2017sparc2, ghosh2017sparc1, ghosh2016higher, suryanarayana2014augmented}. In the present work, we extend M-SPARC to include a spin-constrained formalism that enables calculations at fixed total magnetization. This capability has been implemented for both orbital-free and Kohn--Sham DFT and is used here to evaluate the curvature of the total energy with respect to magnetization, which forms the basis of the magnetic stability analysis below.

We investigate a representative set of metallic systems spanning paramagnetic and ferromagnetic behavior. The paramagnetic metals considered are Al in the face-centered cubic (fcc) structure, a weak paramagnet, and Pd in the face-centered cubic structure, a strongly enhanced paramagnet. For the ferromagnetic case, we consider the $3d$ transition metals Fe in the body-centered cubic (bcc) structure, Co in the hexagonal close-packed (hcp) structure, and Ni in the face-centered cubic structure. Together, these materials span nearly free-electron and transition-metal regimes and provide a systematic test set for assessing magnetic trends and for evaluating the accuracy of TFW orbital-free DFT.

In this work, three types of calculations are performed: self-consistent orbital-free DFT (OF--OF), self-consistent Kohn--Sham DFT reference calculations (KS--KS), and non-self-consistent Kohn--Sham calculations evaluated using the orbital-free ground-state densities (OF--KS). The OF--KS approach was considered based on previous studies demonstrating its ability to accurately reproduce spin-unpolarized Kohn--Sham DFT results \cite{ullmo2001semiclassical,10.1063/1.2176610,10.1063/1.2821101,10.1063/5.0146167}. Lattice constants were set to their experimental values \cite{LandoltBornstein1994:sm_lbs_978-3-540-47399-2_6}. The cells considered contain one atom for the fcc and bcc structures and two atoms for the hcp structure. Exchange--correlation effects were treated within the local spin-density approximation (LSDA) using the Perdew--Wang parameterization \cite{perdew1992}.

Orbital-free calculations employ the PGBRV0.2 local pseudopotentials \cite{rios2025pseudopotentials}, designed for compatibility with the TFW kinetic-energy functional, while the self-consistent Kohn--Sham reference calculations use optimized norm-conserving Vanderbilt (ONCV) pseudopotentials \cite{hamann2013} from the Shojaei--Pask--Medford--Suryanarayana (SPMS) set \cite{SHOJAEI2023108594}. The non-self-consistent Kohn--Sham calculations employ the same pseudopotentials as the orbital-free calculations to ensure consistency with the underlying orbital-free densities. For Al, the Goodwin--Needs--Heine local pseudopotential is employed in all calculations \cite{goodwin1990pseudopotential}, consistent with previous validation studies \cite{10.1063/5.0146167,Thapa2025}. Brillouin-zone integrations for Kohn--Sham calculations employed a $13\times13\times13$ grid for the cubic structures and a $13\times13\times8$ grid for the hcp structure. The real-space discretization used a grid spacing of $\Delta x = 0.25$~bohr together with twelfth-order finite differences, which converges the total energies to within $10^{-4}$~ha/atom. The calculations were performed using the fixed-spin-moment (constrained magnetization) approach described above.

\subsection{Paramagnets: Al and Pd}

\begin{figure}[h]
\centering

\subfloat[Al \label{fig:AlLandau}]{
\includegraphics[width=0.3\textwidth]{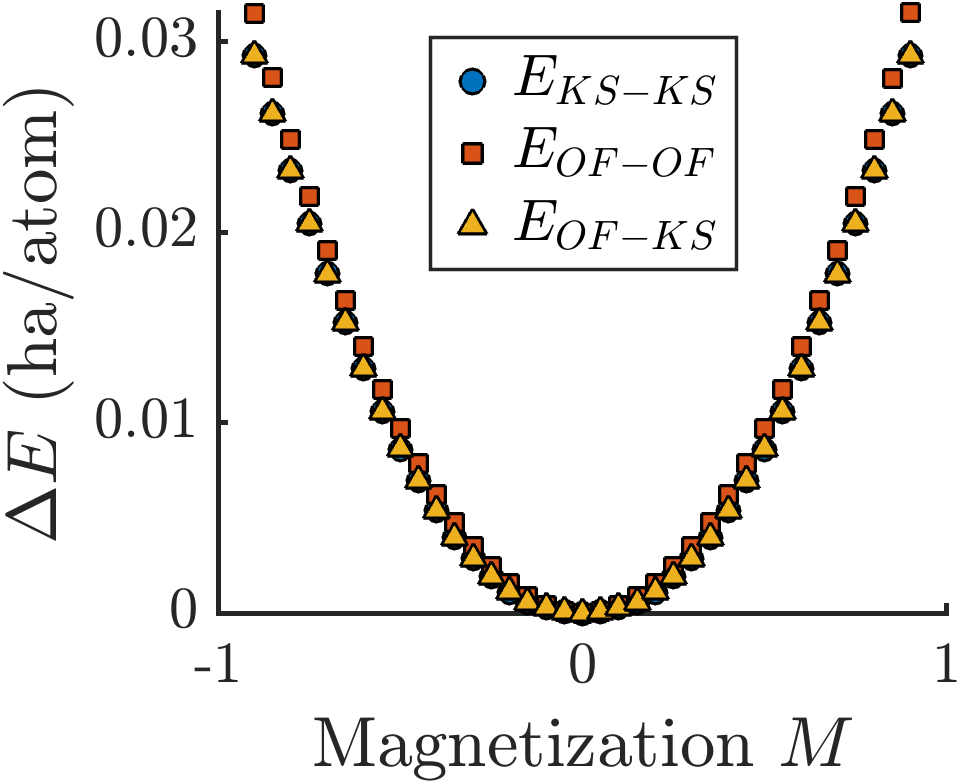}
}

\vspace{0.6em}

\subfloat[Pd \label{fig:PdLandau}]{
\includegraphics[width=0.3\textwidth]{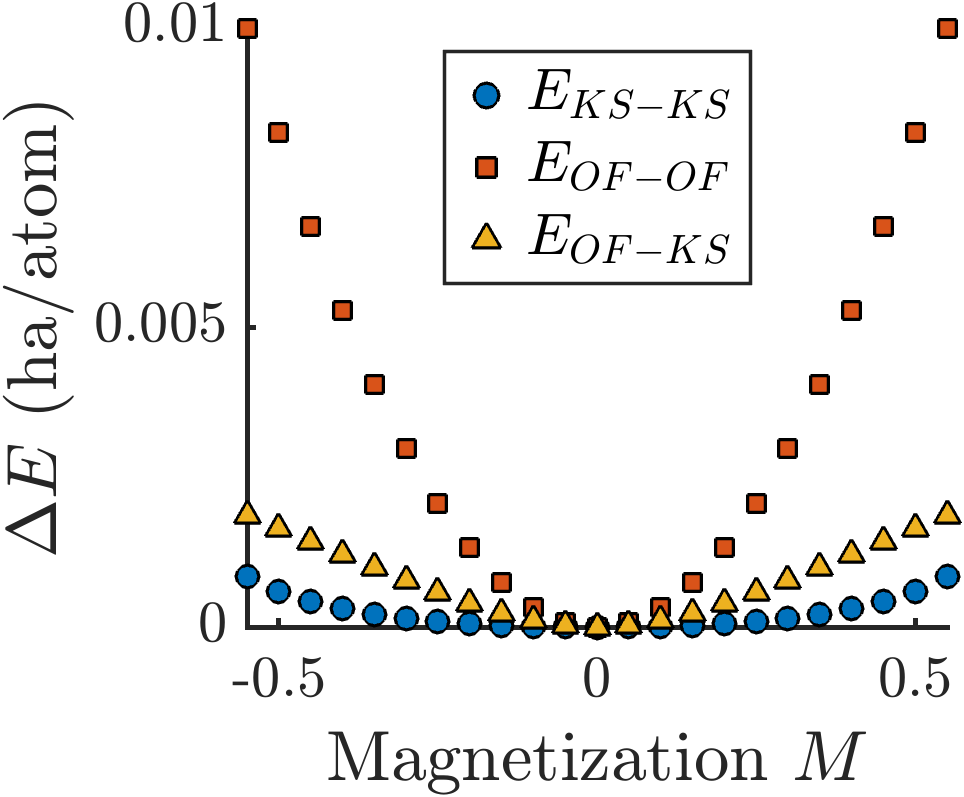}
}

\caption{Variation of the total energy relative to the paramagnetic state as a function of magnetization for the paramagnetic metals (a) Al and (b) Pd.}
\label{fig:LandauParamagnetism}

\end{figure}

Al and Pd serve as reference paramagnetic systems for validating the predicted magnetic response. Al is a weak, nearly free-electron paramagnet that lies far from magnetic instability, whereas Pd is a strongly enhanced paramagnet that is close to the ferromagnetic threshold. These systems therefore provide complementary regimes for assessing the magnetic stability analysis. Figure~\ref{fig:LandauParamagnetism} shows the variation of the total energy with respect to the total magnetization for the three methods. Table~\ref{tab:chi_al_pd} reports the inverse susceptibility $\chi^{-1}$ extracted from the curvature of the energy profile at $M=0$, along with the deviation from the Kohn--Sham benchmark $\Delta\chi^{-1}$. 

\begin{table}[t]

\centering
\setlength{\tabcolsep}{5.0pt}
\renewcommand{\arraystretch}{1.2}

\caption{\justifying Inverse magnetic susceptibility $\chi^{-1}$ and its deviation from the Kohn--Sham benchmark $\Delta\chi^{-1}$ for the paramagnetic metals Al and Pd.}

\label{tab:chi_al_pd}

\begin{tabular}{l l c c}

\toprule
System & Method & $\chi^{-1}\,(\mathrm{ha/atom})$ & $\Delta\chi^{-1}\,(\mathrm{ha/atom})$ \\

\midrule
\multirow{3}{*}{\textbf{Al}}

& KS--KS & $0.0597$  & -- \\
& OF--OF & $0.0775$  & $0.0178$ \\
& OF--KS & $0.0601$  & $0.0004$ \\

\midrule
\multirow{3}{*}{\textbf{Pd}}

& KS--KS & $0.00271$ & -- \\
& OF--OF & $0.0660$  & $0.06329$ \\
& OF--KS & $0.0198$  & $0.01709$ \\

\bottomrule

\end{tabular}

\end{table}

For Al, all three methods yield $\chi^{-1} > 0$, correctly indicating a stable paramagnetic ground state. The differences are therefore quantitative, and relatively small, rather than qualitative. Compared to the KS--KS benchmark, OF--OF underestimates the susceptibility and correspondingly overestimates the magnetic curvature. In contrast, the mixed OF--KS scheme almost completely restores the benchmark response. This behavior is evident in Fig.~\ref{fig:LandauParamagnetism}(a), where the OF--KS curve closely follows the KS--KS profile near $M=0$, while the OF--OF curve remains noticeably steeper. Consistently, the deviation in inverse susceptibility is reduced from $0.0178$ in OF--OF to only $0.0004$ in OF--KS, indicating that for Al the mixed approach reproduces the magnetic response essentially at the KS benchmark level.

For Pd, all three methods again give $\chi^{-1} > 0$, indicating that the paramagnetic state remains formally stable. However, the KS--KS value $\chi^{-1}=0.00271$ is extremely small, implying a very shallow curvature of the magnetic energy surface and therefore a system that lies close to ferromagnetic instability. This near-critical behavior is evident in Fig.~\ref{fig:LandauParamagnetism}(b), where the KS--KS energy profile is nearly flat around $M=0$. In contrast, OF--OF predicts a much larger inverse susceptibility and therefore a significantly steeper energy curve, failing to capture the delicate magnetic response of Pd. The OF--KS scheme moves the result in the correct direction and reduces the deviation $\Delta\chi^{-1}$ from $0.06329$  to $0.01709$, although the recovery remains only partial. Overall, the mixed OF--KS approach substantially improves the predicted magnetic response relative to fully self-consistent orbital-free calculations.

\subsection{Ferromagnets: Fe, Co, and Ni}

\begin{figure*}[t]
\centering
\setlength{\tabcolsep}{4pt}
\begin{tabular}{ccc}
\subfloat[Fe\label{fig:Fe}]{
\includegraphics[width=0.30\textwidth]{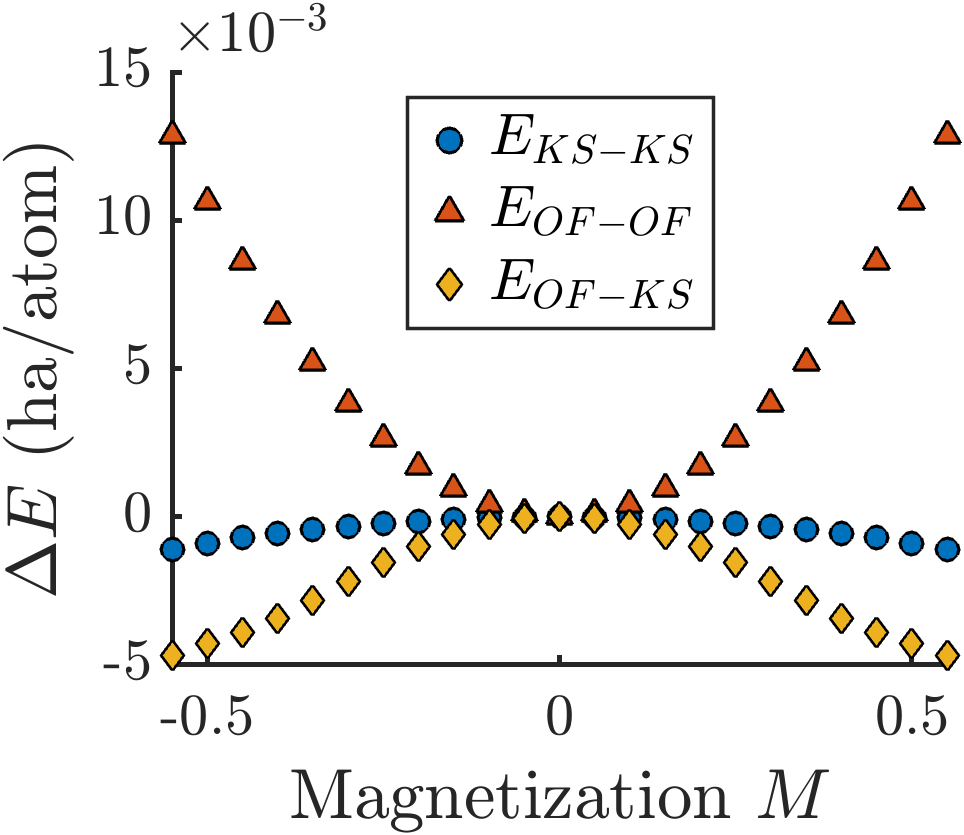}
} &
\subfloat[Co\label{fig:Co}]{
\includegraphics[width=0.30\textwidth]{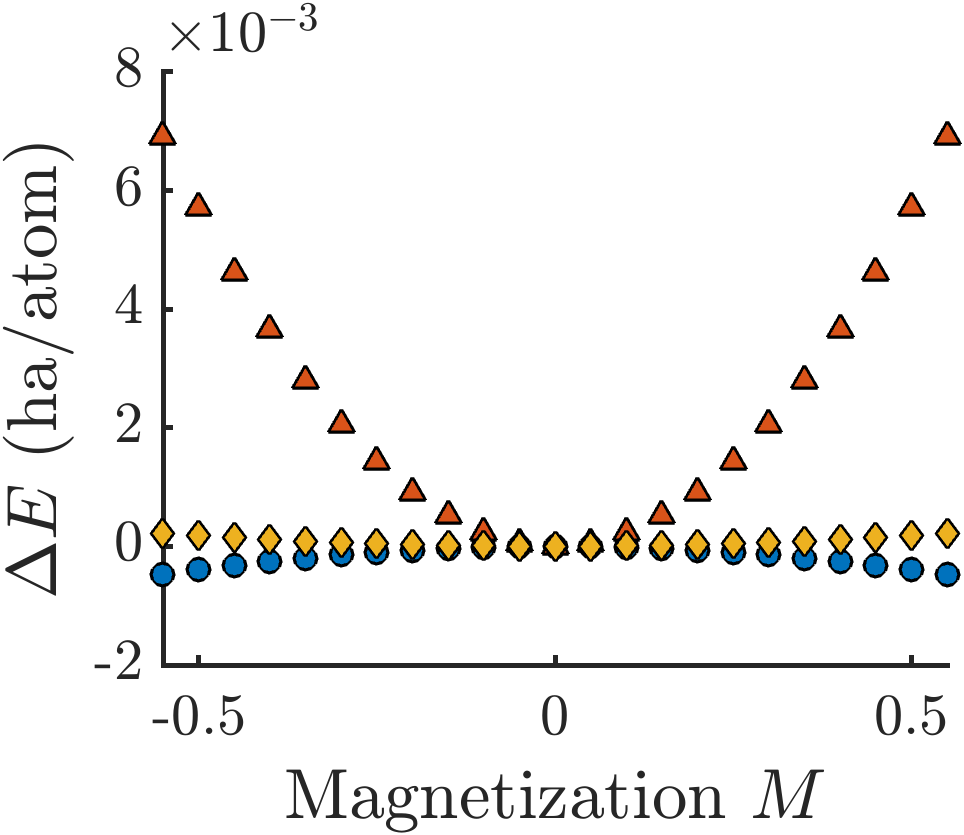}
} &
\subfloat[Ni\label{fig:Ni}]{
\includegraphics[width=0.312\textwidth]{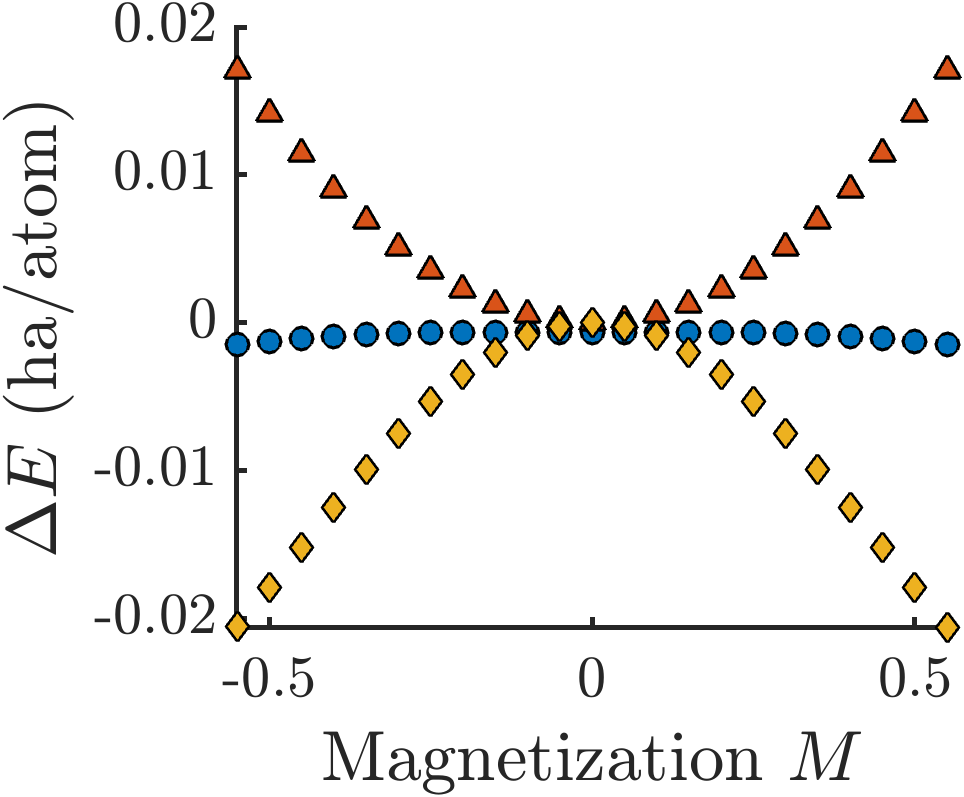}
}
\end{tabular}

\caption{Variation of the total energy relative to the paramagnetic state as a function of magnetization for the ferromagnetic metals (a) Fe, (b) Co, and (c) Ni.}
\label{fig:LandauAll}
\end{figure*}

The $3d$ transition metals Fe, Co, and Ni provide representative ferromagnetic systems in which magnetism arises from exchange splitting of partially filled $d$ bands. Figure~\ref{fig:LandauAll} shows the variation of the total energy with respect to magnetization, while Table~\ref{tab:chi_feconi} reports the corresponding inverse susceptibilities $\chi^{-1}$ and their deviation from the Kohn--Sham benchmark $\Delta\chi^{-1}$.

\begin{table}[h]
\centering
\setlength{\tabcolsep}{9pt}
\renewcommand{\arraystretch}{1.2}

\caption{\justifying Inverse magnetic susceptibility $\chi^{-1}$ and its deviation from the Kohn--Sham benchmark $\Delta\chi^{-1}$ for the ferromagnetic metals Fe, Co, and Ni.}

\begin{tabular}{l l c c}
\toprule
System & Method & $\chi^{-1}\,(\mathrm{ha/atom})$ & $\Delta\chi^{-1}\,(\mathrm{ha/atom})$ \\
\midrule

\multirow{3}{*}{\textbf{Fe}}
& KS--KS & $-0.00794$ & -- \\
& OF--OF & $0.08489$  & $0.09283$ \\
& OF--KS & $-0.06868$ & $-0.06074$ \\

\midrule
\multirow{3}{*}{\textbf{Co}}
& KS--KS & $-0.00314$ & -- \\
& OF--OF & $0.04568$  & $0.04882$ \\
& OF--KS & $0.00142$  & $0.00456$ \\

\midrule
\multirow{3}{*}{\textbf{Ni}}
& KS--KS & $-0.00595$ & -- \\
& OF--OF & $0.11274$  & $0.11869$ \\
& OF--KS & $-0.18018$ & $-0.17423$ \\

\bottomrule
\end{tabular}

\label{tab:chi_feconi}
\end{table}

For all three materials, the KS--KS calculations give $\chi^{-1} < 0$, consistent with the instability of the paramagnetic state toward ferromagnetism. In contrast, OF--OF calculations yield $\chi^{-1} > 0$ for Fe, Co, and Ni, incorrectly predicting a stable paramagnetic state in each case. This failure is reflected in the large positive deviations $\Delta\chi^{-1}$ relative to the KS benchmark, indicating a substantial overestimation of the magnetic curvature.

The mixed OF--KS scheme significantly improves the predicted magnetic response. For Fe and Ni, the sign of $\chi^{-1}$ becomes negative again, restoring the correct qualitative behavior and bringing the results much closer to the KS reference.  For Co, OF--KS likewise reduces the inverse susceptibility substantially---from $0.04568$ in OF--OF to $0.00142$---placing the system very close to the instability threshold, although the sign remains slightly positive. The corresponding energy profiles in Fig.~\ref{fig:LandauAll} show that the OF--KS curves follow the KS--KS curvature much more closely near $M=0$, whereas the OF--OF curves remain significantly steeper.

Taken together, these results highlight important limitations of TFW OF-DFT for itinerant magnetism. The failure of self-consistent OF--DFT to reproduce ferromagnetism in Fe, Co, and Ni can be attributed to the pronounced variation of the electronic density of states near the Fermi level. These materials exhibit narrow $d$-band peaks at $E_F$, making the magnetic response highly sensitive to the curvature of the energy functional. The TFW kinetic-energy functional does not accurately capture this curvature in the presence of such sharp electronic features, leading to an incorrect magnetic response.

The partial success of the OF--KS scheme, which recovers ferromagnetism for Fe and Ni and yields a near-critical response for Co, indicates that the orbital-free densities retain sufficient fidelity to support perturbative corrections. This observation suggests a possible pathway for improving orbital-free-based magnetic predictions through hybrid approaches that combine orbital-free densities with Kohn--Sham energy evaluations, particularly for describing linear response in systems with strong electronic-structure variations near the Fermi level.



\section{Concluding remarks}

We assessed the ability of the TFW functional within orbital-free DFT to describe itinerant magnetism. Magnetic stability was evaluated from the susceptibility obtained from the second derivative of the total energy with respect to the net magnetization. Calculations were carried out for the paramagnetic metals Al and Pd and the canonical ferromagnets Fe, Co, and Ni, and the results were benchmarked against Kohn-Sham DFT. The orbital-free calculations showed poor agreement with the Kohn-Sham predictions, particularly for the transition metals where the magnetic response is governed by narrow $d$-band features near the Fermi level. These sharp electronic-structure features make the magnetic energy curvature highly sensitive to the accuracy of the kinetic-energy functional. Using the orbital-free DFT ground-state densities in a non-self-consistent KS evaluation improved the qualitative behavior and partially recovered the correct magnetic trends, although quantitative agreement remained limited. 

Overall, the results highlight fundamental limitations of the TFW functional for describing itinerant magnetism. Future work will explore whether more advanced orbital-free kinetic-energy functionals can better capture the magnetic response of itinerant systems and improve the predictive capability of orbital-free DFT for magnetic materials.

\begin{acknowledgments}
P.S. gratefully acknowledges the support of the U.S. Department of Energy, Office of Science under grant DE-SC0023445.
\end{acknowledgments}

\bibliography{bib}

@article{gunnarsson1976band,
  title   = {Band model for magnetism of transition metals in the spin-density-functional formalism},
  author  = {Gunnarsson, O.},
  journal = {J. Phys. F: Met. Phys.},
  volume  = {6},
  number  = {4},
  pages   = {587},
  year    = {1976},
  doi     = {10.1088/0305-4608/6/4/018}
}

@article{doi:10.1021/acs.chemrev.2c00758,
author = {Mi, Wenhui and Luo, Kai and Trickey, S. B. and Pavanello, Michele},
title = {Orbital-Free Density Functional Theory: An Attractive Electronic Structure Method for Large-Scale First-Principles Simulations},
journal = {Chemical Reviews},
volume = {123},
number = {21},
pages = {12039-12104},
year = {2023},
doi = {10.1021/acs.chemrev.2c00758},
}

@article{oliver1979spin,
  title={Spin-density gradient expansion for the kinetic energy},
  author={Oliver, GL and Perdew, JP},
  journal={Physical Review A},
  volume={20},
  number={2},
  pages={397},
  year={1979},
  publisher={APS},
  doi = {https://doi.org/10.1103/PhysRevA.20.397}
}

@article{janas2023enhancing,
  title={Enhancing electron correlation at a 3d ferromagnetic surface},
  author={Janas, David Maximilian and Droghetti, Andrea and Ponzoni, Stefano and Cojocariu, Iulia and Jugovac, Matteo and Feyer, Vitaliy and Radonji{\'c}, Milo{\v{s}} M and Rungger, Ivan and Chioncel, Liviu and Zamborlini, Giovanni and others},
  journal={Advanced Materials},
  volume={35},
  number={3},
  pages={2205698},
  year={2023},
  publisher={Wiley Online Library},
  doi = {https://doi.org/10.1002/adma.202205698}
}

@article{perdew1992,
  author  = {J. P. Perdew and Y. Wang},
  title   = {Accurate and simple analytic representation of the electron-gas correlation energy},
  journal = {Phys. Rev. B},
  volume  = {45},
  pages   = {13244--13249},
  year    = {1992},
  doi     = {10.1103/PhysRevB.45.13244}
}

@article{rios2025pseudopotentials,
  author  = {Rios-Vargas, Valeria and Oyeniyi, Ezekiel and Shao, Xuecheng and Elsayed, Wala Fathelrahman Ibrahim and Ogenyi, Sunday Joseph and Okello, Alex and Pavanello, Michele},
  title   = {Pseudopotentials for Orbital-Free DFT: Capturing Nonlocality and Correcting Functional Approximants},
  journal = {arXiv},
  volume  = {2511.19892},
  year    = {2025},
  doi     = {10.48550/arXiv.2511.19892}
}

@article{hamann2013,
  author  = {D. R. Hamann},
  title   = {Optimized norm-conserving Vanderbilt pseudopotentials},
  journal = {Phys. Rev. B},
  volume  = {88},
  pages   = {085117},
  year    = {2013},
  doi     = {10.1103/PhysRevB.88.085117}
}

@article{goodwin1990pseudopotential,
  title={A pseudopotential total energy study of impurity-promoted intergranular embrittlement},
  author={Goodwin, L and Needs, RJ and Heine, Volker},
  journal={Journal of Physics: Condensed Matter},
  volume={2},
  number={2},
  pages={351},
  year={1990},
  publisher={IOP Publishing},
  doi = {10.1088/0953-8984/2/2/011}
}

@article{PhysRev.136.B864,
  title = {Inhomogeneous Electron Gas},
  author = {Hohenberg, P. and Kohn, W.},
  journal = {Phys. Rev.},
  volume = {136},
  issue = {3B},
  pages = {B864--B871},
  numpages = {0},
  year = {1964},
  month = {Nov},
  publisher = {American Physical Society},
  doi = {10.1103/PhysRev.136.B864},
  url = {https://link.aps.org/doi/10.1103/PhysRev.136.B864}
}

@article{PhysRev.140.A1133,
  title = {Self-Consistent Equations Including Exchange and Correlation Effects},
  author = {Kohn, W. and Sham, L. J.},
  journal = {Phys. Rev.},
  volume = {140},
  issue = {4A},
  pages = {A1133--A1138},
  numpages = {0},
  year = {1965},
  month = {Nov},
  publisher = {American Physical Society},
  doi = {10.1103/PhysRev.140.A1133},
  url = {https://link.aps.org/doi/10.1103/PhysRev.140.A1133}
}

@article{PhysRev.75.1561,
  title = {Equations of State of Elements Based on the Generalized {F}ermi-{T}homas Theory},
  author = {Feynman, R. P. and Metropolis, N. and Teller, E.},
  journal = {Phys. Rev.},
  volume = {75},
  issue = {10},
  pages = {1561--1573},
  numpages = {0},
  year = {1949},
  month = {May},
  publisher = {American Physical Society},
  doi = {10.1103/PhysRev.75.1561},
  url = {https://link.aps.org/doi/10.1103/PhysRev.75.1561}
}

@article{PhysRevA.20.586,
  title = {Gradient correction to the statistical electronic free energy at nonzero temperatures: Application to equation-of-state calculations},
  author = {Perrot, F.},
  journal = {Phys. Rev. A},
  volume = {20},
  issue = {2},
  pages = {586--594},
  numpages = {0},
  year = {1979},
  month = {Aug},
  publisher = {American Physical Society},
  doi = {10.1103/PhysRevA.20.586},
  url = {https://link.aps.org/doi/10.1103/PhysRevA.20.586}
}

@article{Thomas_1927, title={The calculation of atomic fields}, volume={23}, DOI={10.1017/S0305004100011683}, number={5}, journal={Mathematical Proceedings of the Cambridge Philosophical Society}, author={Thomas, L. H.}, year={1927}, pages={542–548}}

@article{benguria1981thomas,
  author  = {Rafael Benguria and Haim Brezis and Elliott H. Lieb},
  title   = {The {T}homas-{F}ermi-von {W}eizs{\"a}cker theory of atoms and molecules},
  journal = {Communications in Mathematical Physics},
  volume  = {79},
  number  = {2},
  pages   = {167--180},
  year    = {1981},
  doi     = {10.1007/BF01942059},
  publisher = {Springer}
}

@article{xia2012can,
  author  = {Xia, Junchao and Huang, Chen and Shin, Ilgyou and Carter, Emily A.},
  title   = {Can orbital-free density functional theory simulate molecules?},
  journal = {J. Chem. Phys.},
  volume  = {136},
  number  = {8},
  pages   = {084102},
  year    = {2012},
  month   = feb,
  doi     = {10.1063/1.3685604}
}

@article{PhysRevB.88.161108,
  title = {Nonempirical generalized gradient approximation free-energy functional for orbital-free simulations},
  author = {Karasiev, Valentin V. and Chakraborty, Debajit and Shukruto, Olga A. and Trickey, S. B.},
  journal = {Phys. Rev. B},
  volume = {88},
  issue = {16},
  pages = {161108},
  numpages = {5},
  year = {2013},
  month = {Oct},
  publisher = {American Physical Society},
  doi = {10.1103/PhysRevB.88.161108},
  url = {https://link.aps.org/doi/10.1103/PhysRevB.88.161108}
}

@article{PhysRevB.98.041111,
  title = {A simple generalized gradient approximation for the noninteracting kinetic energy density functional},
  author = {Luo, Kai and Karasiev, Valentin V. and Trickey, S. B.},
  journal = {Phys. Rev. B},
  volume = {98},
  issue = {4},
  pages = {041111},
  numpages = {5},
  year = {2018},
  month = {Jul},
  publisher = {American Physical Society},
  doi = {10.1103/PhysRevB.98.041111},
  url = {https://link.aps.org/doi/10.1103/PhysRevB.98.041111}
}

@article{PhysRevB.100.165111,
  title = {Semilocal kinetic energy functionals with parameters from neutral atoms},
  author = {Lehtom\"aki, Jouko and Lopez-Acevedo, Olga},
  journal = {Phys. Rev. B},
  volume = {100},
  issue = {16},
  pages = {165111},
  numpages = {9},
  year = {2019},
  month = {Oct},
  publisher = {American Physical Society},
  doi = {10.1103/PhysRevB.100.165111},
  url = {https://link.aps.org/doi/10.1103/PhysRevB.100.165111}
}

@article{Laricchia2011,
  author  = {Laricchia, S. and Fabiano, E. and Constantin, L. A. and Della Sala, F.},
  title   = {Generalized Gradient Approximations of the Noninteracting Kinetic Energy from the Semiclassical Atom Theory: Rationalization of the Accuracy of the Frozen Density Embedding Theory for Nonbonded Interactions},
  journal = {J. Chem. Theory Comput.},
  year    = {2011},
  volume  = {7},
  number  = {8},
  pages   = {2439--2451},
  doi     = {https://doi.org/10.1021/ct200382w}
}

@article{PhysRevB.99.155137,
  title = {Semilocal properties of the {P}auli kinetic potential},
  author = {Constantin, Lucian A.},
  journal = {Phys. Rev. B},
  volume = {99},
  issue = {15},
  pages = {155137},
  numpages = {12},
  year = {2019},
  month = {Apr},
  publisher = {American Physical Society},
  doi = {10.1103/PhysRevB.99.155137},
  url = {https://link.aps.org/doi/10.1103/PhysRevB.99.155137}
}

@article{PhysRevB.32.7868,
  title = {Nonlocal kinetic energy functional for nonhomogeneous electron systems},
  author = {Chac\'on, E. and Alvarellos, J. E. and Tarazona, P.},
  journal = {Phys. Rev. B},
  volume = {32},
  issue = {12},
  pages = {7868--7877},
  numpages = {0},
  year = {1985},
  month = {Dec},
  publisher = {American Physical Society},
  doi = {10.1103/PhysRevB.32.7868},
  url = {https://link.aps.org/doi/10.1103/PhysRevB.32.7868}
}

@article{PhysRevB.53.9509,
  title = {Nonlocal kinetic-energy-density functionals},
  author = {Garc\'{\i}a-Gonz\'alez, P. and Alvarellos, J. E. and Chac\'on, E.},
  journal = {Phys. Rev. B},
  volume = {53},
  issue = {15},
  pages = {9509--9512},
  numpages = {0},
  year = {1996},
  month = {Apr},
  publisher = {American Physical Society},
  doi = {10.1103/PhysRevB.53.9509},
  url = {https://link.aps.org/doi/10.1103/PhysRevB.53.9509}
}

@article{PhysRevB.57.4857,
  title = {Nonlocal symmetrized kinetic-energy density functional: Application to simple surfaces},
  author = {Garc\'{\i}a-Gonz\'alez, P. and Alvarellos, J. E. and Chac\'on, E.},
  journal = {Phys. Rev. B},
  volume = {57},
  issue = {8},
  pages = {4857--4862},
  numpages = {0},
  year = {1998},
  month = {Feb},
  publisher = {American Physical Society},
  doi = {10.1103/PhysRevB.57.4857},
  url = {https://link.aps.org/doi/10.1103/PhysRevB.57.4857}
}

@article{PhysRevB.45.13196,
  title = {Kinetic-energy functional of the electron density},
  author = {Wang, Lin-Wang and Teter, Michael P.},
  journal = {Phys. Rev. B},
  volume = {45},
  issue = {23},
  pages = {13196--13220},
  numpages = {0},
  year = {1992},
  month = {Jun},
  publisher = {American Physical Society},
  doi = {10.1103/PhysRevB.45.13196},
  url = {https://link.aps.org/doi/10.1103/PhysRevB.45.13196}
}

@article{PhysRevB.60.16350,
  title = {Orbital-free kinetic-energy density functionals with a density-dependent kernel},
  author = {Wang, Yan Alexander and Govind, Niranjan and Carter, Emily A.},
  journal = {Phys. Rev. B},
  volume = {60},
  issue = {24},
  pages = {16350--16358},
  numpages = {0},
  year = {1999},
  month = {Dec},
  publisher = {American Physical Society},
  doi = {10.1103/PhysRevB.60.16350},
  url = {https://link.aps.org/doi/10.1103/PhysRevB.60.16350}
}

@article{SHOJAEI2023108594,
title = {Soft and transferable pseudopotentials from multi-objective optimization},
journal = {Computer Physics Communications},
volume = {283},
pages = {108594},
year = {2023},
issn = {0010-4655},
doi = {https://doi.org/10.1016/j.cpc.2022.108594},
url = {https://www.sciencedirect.com/science/article/pii/S0010465522003137},
author = {Mostafa Faghih Shojaei and John E. Pask and Andrew J. Medford and Phanish Suryanarayana}
}

@article{10.1063/5.0146167,
    author = {Thapa, Bishal and Jing, Xin and Pask, John E. and Suryanarayana, Phanish and Mazin, Igor I.},
    title = {Assessing the source of error in the {T}homas–{F}ermi–von {W}eizsäcker density functional},
    journal = {The Journal of Chemical Physics},
    volume = {158},
    number = {21},
    pages = {214112},
    year = {2023},
    month = {06},
    issn = {0021-9606},
    doi = {10.1063/5.0146167},
    url = {https://doi.org/10.1063/5.0146167},
}

@article{Thapa2025,
  author  = {Thapa, Bishal and Oellerich, Tracey G. and Emelianenko, Maria and Suryanarayana, Phanish and Mazin, Igor I.},
  title   = {Orbital-free density functionals based on real and reciprocal space separation},
  journal = {npj Comput. Mater.},
  year    = {2025},
  volume  = {11},
  number  = {1},
  pages   = {149},
  doi     = {10.1038/s41524-025-01643-0}
}

@article{xu2020m,
  author    = {Qimen Xu and Abhiraj Sharma and Phanish Suryanarayana},
  title     = {{M-SPARC}: Matlab-simulation package for ab-initio real-space calculations},
  journal   = {SoftwareX},
  volume    = {11},
  pages     = {100423},
  year      = {2020},
  doi       = {10.1016/j.softx.2020.100423},
  publisher = {Elsevier}
}

@article{xu2021sparc,
  title={{SPARC}: Simulation package for ab-initio real-space calculations},
  author={Xu, Qimen and Sharma, Abhiraj and Comer, Benjamin and Huang, Hua and Chow, Edmond and Medford, Andrew J and Pask, John E and Suryanarayana, Phanish},
  journal={SoftwareX},
  volume={15},
  pages={100709},
  year={2021},
  publisher={Elsevier},
  doi = {https://doi.org/10.1016/j.softx.2021.100709}
}

@Article{	  ghosh2017sparc1,
  title		= {{SPARC}: Accurate and efficient finite-difference
		  formulation and parallel implementation of Density
		  Functional Theory: Isolated clusters},
  author	= {Swarnava Ghosh and Phanish Suryanarayana},
  journal	= {Computer Physics Communications},
  volume	= {212},
  pages		= {189--204},
  year		= {2017},
  publisher	= {Elsevier},
  doi = {https://doi.org/10.1016/j.cpc.2016.09.020}
}

@Article{	  ghosh2017sparc2,
  title		= {{SPARC}: Accurate and efficient finite-difference
		  formulation and parallel implementation of Density
		  Functional Theory: Extended systems},
  author	= {Swarnava Ghosh and Phanish Suryanarayana},
  journal	= {Computer Physics Communications},
  volume	= {216},
  pages		= {109--125},
  year		= {2017},
  publisher	= {Elsevier},
  doi = {https://doi.org/10.1016/j.cpc.2017.02.019}
  
}

@article{zhang2023version,
  author    = {Boqin Zhang and Xin Jing and Shashikant Kumar and Phanish Suryanarayana},
  title     = {Version 2.0.0-{{M-SPARC}}: {Matlab}-simulation package for ab-initio real-space calculations},
  journal   = {SoftwareX},
  volume    = {21},
  pages     = {101295},
  year      = {2023},
  doi       = {10.1016/j.softx.2022.101295},
  publisher = {Elsevier}
}

@article{ghosh2016higher,
  title={Higher-order finite-difference formulation of periodic orbital-free density functional theory},
  author={Ghosh, Swarnava and Suryanarayana, Phanish},
  journal={Journal of Computational Physics},
  volume={307},
  pages={634--652},
  year={2016},
  publisher={Elsevier},
  doi = {https://doi.org/10.1016/j.jcp.2015.12.027}
}

@article{suryanarayana2014augmented,
  title={Augmented Lagrangian formulation of orbital-free density functional theory},
  author={Suryanarayana, Phanish and Phanish, Deepa},
  journal={Journal of Computational Physics},
  volume={275},
  pages={524--538},
  year={2014},
  publisher={Elsevier},
  doi = {https://doi.org/10.1016/j.jcp.2014.07.006}
}

@misc{LandoltBornstein1994:sm_lbs_978-3-540-47399-2_6,
  author    = {Chiarotti, G.},
  title     = {Crystal structures and bulk lattice parameters of materials},
  booktitle = {Landolt-B{\"o}rnstein -- Group III Condensed Matter, Volume 24B: Electronic and Vibrational Properties},
  publisher = {Springer-Verlag},
  year      = {1994},
  doi       = {10.1007/10086058_6},
  url       = {https://materials.springer.com/lb/docs/sm_lbs_978-3-540-47399-2_6}
}

@article{ullmo2001semiclassical,
  title={Semiclassical density functional theory: Strutinsky energy corrections in quantum dots},
  author={Ullmo, Denis and Nagano, Tatsuro and Tomsovic, Steven and Baranger, Harold U},
  journal={Physical Review B},
  volume={63},
  number={12},
  pages={125339},
  year={2001},
  publisher={APS},
  doi = {https://doi.org/10.1103/PhysRevB.63.125339}
}

@article{10.1063/1.2176610,
  author  = {Zhou, Baojing and Wang, Yan Alexander},
  title   = {Orbital-corrected orbital-free density functional theory},
  journal = {J. Chem. Phys.},
  volume  = {124},
  number  = {8},
  pages   = {081107},
  year    = {2006},
  doi     = {10.1063/1.2176610}
}

@article{10.1063/1.2821101,
  author  = {Zhou, Baojing and Wang, Yan Alexander},
  title   = {Accelerating the convergence of the total energy evaluation in density functional theory calculations},
  journal = {J. Chem. Phys.},
  volume  = {128},
  number  = {8},
  pages   = {084101},
  year    = {2008},
  doi     = {10.1063/1.2821101}
}

\end{document}